\documentclass{article}
\usepackage{graphicx} % Required for inserting images
\usepackage{blindtext}
\usepackage{amsmath}
\usepackage[a4paper, total={6in, 8in}]{geometry}
\usepackage{amsfonts}
\usepackage{hyperref} 
\usepackage{xcolor}
\usepackage{scalerel}[2016/12/29]
\usepackage{float}
\usepackage{authblk}

\begin{document}

\title{Perturbative power series for block diagonalisation of Hermitian matrices}
\author[a]{Ishan N. H. Mankodi}
\author[b,c]{David P. DiVincenzo}

\affil[a]{Department of Physics, IIT Madras, Chennai 600036, India.}

\affil[b]{Institute for Quantum Information, RWTH Aachen University, 52056 Aachen, Germany.}

\affil[c]{Peter Gr\"unberg Institute, Theoretical Nanoelectronics, Forschungszentrum J\"ulich, 52425 J\"ulich, Germany.}

%\date{March 2024}

\maketitle

\begin{abstract}
Block diagonalisation of matrices by canonical transformation is important in various fields of physics. Such diagonalization is currently of interest in condensed matter physics, for modelling of gates in superconducting circuits and for studying isolated quantum many-body systems. While the block diagonalisation of a particular Hermitian matrix is not unique, it can be made unique with certain auxiliary conditions. It has been assumed in some recent literature that two of these conditions, ``least action" vs. block-off-diagonality of the generator, lead to identical transformations. We show that this is not the case, and that these two approaches diverge at third order in the small parameter.  
We derive the perturbative power series of the ``least action", exhibiting explicitly the loss of block-off-diagnoality.
\end{abstract}

%Some references not to lose: \cite{doi:10.1142/S0217979203020144},\cite{Jung2020}

%\fading{\tiny Points:
%\begin{itemize}
%    \item{block diagonalization: N excitation sectors (H{\"o}rmann-Schmidt notation)}
%    \item{N=2 is "Schrieffer Wolff" (low vs. high)}
%    \item{objective is always block diagonalization}
%    \item{SW is clearly most used, formalism well known}
%    \item{unitary, generator of unitary}
%    \item{in SW U being least action and S being off-block diagonal are the same}
%    \item{literature has been unclear on whether they are the same}
%    \item{least action is well established: exact formula, series technique (Takahashi, HS)}
%    \item{series is clear, but cumbersome to construct (Takahashi, Kato)}
%\end{itemize}}

\section{Introduction}

There are many applications in which it is desired to apply a unitary transformation $T$ to a Hermitian matrix $H$:
\begin{equation}
H_{block} = T^ {\dagger} H T, \label{blockdi}   
\end{equation}
such that the resulting matrix $H_{block}$ has some specified block-diagonal structure. In physical applications in quantum mechanics, the block structure may involve sectors with a certain excitation number, or it may just correspond to bands of energy. Often a simpler physical model is represented by the part of the block matrix in the sector of interest.

Block diagonalization is obviously not uniquely defined --- one can see this by noting that full diagonalization is a special instance of block diagonalization. Thus, some additional condition must be imposed to make the transformation unique.

An important insight of Cederbaum {\em et al.}~\cite{L.S.Cederbaum_1989} was that a physically relevant and implementable constraint is to require that the block-diagonalizing unitary operator be as close to possible to the identity $I$, in the norm sense. They showed that requiring $||T-I||$ to be minimal leads to a formula for $T$:
\begin{equation}
    T = X\, \scaleobj{0.9}{\mathcal{B}}(X^{\dagger})(\scaleobj{0.9}{\mathcal{B}}(X)\scaleobj{0.9}{\mathcal{B}}(X^{\dagger}))^{-\frac{1}{2}},
    \label{ceder2}
\end{equation}
where, following the notation of \cite{PhysRevA.101.052308}, $X$ is the unitary eigenvector matrix of the hermitian matrix $H$. Equation \eqref{ceder2} introduces the superoperator $\scaleobj{0.9}{\mathcal{B}}(\cdot)$, which sets the block off-diagonal elements of its argument to zero and leaves the block diagonal parts unchanged. That is, $\scaleobj{0.9}{\mathcal{B}}(X)$, acting on square matrix $X$, gives a matrix of the same dimension with matrix elements
\begin{eqnarray}
    \scaleobj{0.9}{\mathcal{B}}(X)_{ij}&=&X_{ij},\,\,\mbox{$i$ and $j$ in the same block,}\nonumber\\
     \scaleobj{0.9}{\mathcal{B}}(X)_{ij}&=&0,\,\,\,\,\mbox{$i$ and $j$ in different blocks.}
\end{eqnarray}
Formula \ref{ceder2} can be found in previous work of Takahashi \cite{MTakahashi_1977}, where the relation to minimal rotation was not remarked; see H{\"o}rmann and Schmidt \cite{10.21468/SciPostPhys.15.3.097} for more discussion.

Note that an immediate consequence of Eq.~(\ref{ceder2}) is a full formula for $H_{block}$:
\begin{equation}
    H_{block}= 
    (\scaleobj{0.9}{\mathcal{B}}(X)\scaleobj{0.9}{\mathcal{B}}(X^{\dagger}))^{-\frac{1}{2}}{\mathcal{B}}(X)\cdot X^{\dagger}HX\cdot
\scaleobj{0.9}{\mathcal{B}}(X^{\dagger})(\scaleobj{0.9}{\mathcal{B}}(X)\scaleobj{0.9}{\mathcal{B}}(X^{\dagger}))^{-\frac{1}{2}}.\label{celder3}
\end{equation}
Here we have highlighted that the structure of this formula is a full diagonalization of $H$, followed by a particular block-diagonal ``back rotation".

The Cederbaum development matched with another specific case that has been generally of very high interest in physical theories, in which the number of blocks is only two, seen at least as far back as in the work of Foldy and Wouthuysen \cite{PhysRev.78.29}. Often the two blocks are  ``low" vs. ``high" energy; for the Foldy-Wouthuysen example, the full ``high energy" theory is the Dirac equation, and the ``low energy sector" is the Schr{\"o}dinger-Pauli equation with relativistic corrections. Such a high-low separation was introduced for the Anderson impurity theory in 1966 by Schrieffer and Wolff \cite{PhysRev.149.491}; they showed that the low energy sector of this theory is the Kondo model. There is a large community that refers to the high-low block diagonalization of any physical theory as a ``Schrieffer-Wolff" transformation.

In Schrieffer-Wolff literature the two-block version of Eq.~\eqref{ceder2} was known, but it has generally been thought of as a theory defined by a series expansion of $T$. In this work, another criterion, separate from the minimal-norm one, was developed to specify $T$. This alternative criterion focuses on the generator $S$ of the unitary transformation $T$, viz., $T=e^{-iS}$.  It was found that a series that is developed based on the constraint of block-off-diagonality of $S$, that is, $\scaleobj{0.9}{\mathcal{B}}(S)=0$, leads to a straightforward derivation of a physically appealing series. It was only later known that the two criteria are in fact equivalent: Assuming $\scaleobj{0.9}{\mathcal{B}}(S)=0$ leads to the series representation for Eq. \eqref{ceder2}.

This is well established for the Schrieffer-Wolff case, that is, the two-block diagonalization. The question of whether this equivalence extends to the case of more than two blocks has not been clearly explored in previous work, even though this prescription $\scaleobj{0.9}{\mathcal{B}}(S)=0$ has been used in other applications of multi-block diagonalization \cite{doi:10.1142/S0217979203020144}. In the recent work of Magesan and Gambetta on superconducting qubits \cite{PhysRevA.101.052308} it is assumed that the two remain equivalent. 

It is the purpose of this short note to show that the two criteria for block diagonalization are {\em not} generally equivalent for the multi-block case. We find that the series representations associated with these two different criteria agree up to second order, but diverge from one another at third order and beyond. We will give the explicit series for the Cederbaum criterion up to this order. We will point out some further open questions that our observations raise.

\section{Formalism}
We consider the case of an unperturbed Hamiltonian $H_0$ (assumed already to have a certain block-diagonal structure) and a perturbation term $H_1$ with control parameter $\lambda$ such that 
\begin{equation}
\label{Hbasic}
    H = H_0 + \lambda H_1.
\end{equation}

We will discuss the series representation of the canonical transformation $T = e^{-i S}$ that transforms Hamiltonian $H$ into an effective block diagonal Hamiltonian $H_{block}$ as per Eq. \eqref{blockdi}. 
%$$H_{block} = T^ {\dagger} H T$$
\\
As discussed above, we impose the ``least action" condition of $\lVert T - I \rVert$ = minimum (this has also been called the ``direct rotation" condition \cite{BRAVYI20112793}). The unitary matrix is then uniquely given by Eq. \eqref{ceder2} \cite{L.S.Cederbaum_1989}. For a series analysis, a choice of the Riemann sheet for the square root function must be made. For cases of a small perturbation (small $\lambda$), $X$ will be close to identity, so all its eigenvalues will be close to $+1$, in which case the square root close to $+1$ is always to be chosen. \cite{BRAVYI20112793} has a careful discussion of what happens if one of these eigenvalues approaches $-1$, as it can when $\lambda$ is large.
%\fading{\begin{equation}
%    T = S\, \scaleobj{0.9}{\mathcal{B}}(S^{\dagger})(\scaleobj{0.9}{\mathcal{B}}(S)\scaleobj{0.9}{\mathcal{B}}(S^{\dagger}))^{-\frac{1}{2}}
    \label{ceder}
%\end{equation}
%where $S$ is the unitary eigenvector matrix of the hermitian matrix $H$. Equation \eqref{ceder} introduces the superoperator $\scaleobj{0.9}{\mathcal{B}}(\cdot)$, which sets the block off-diagonal elements of its argument to zero and leaves the block diagonal parts unchanged. That is, $\scaleobj{0.9}{\mathcal{B}}(S)$, acting on square matrix $S$, gives a matrix of the same dimension with matrix elements
%\begin{eqnarray}
%    \scaleobj{0.9}{\mathcal{B}}(S)_{ij}&=&z_{ij},\,\,\mbox{$i$ and $j$ in the same block,}\nonumber\\
%     \scaleobj{0.9}{\mathcal{B}}(S)_{ij}&=&0,\,\,\,\,\mbox{$i$ and $j$ in different blocks.}
%\end{eqnarray} 
%and $\scaleobj{0.9}{\mathcal{B}}(S)$ is its block diagonal part.}

The eigenvector matrix $X$, the matrix that diagonalises the Hamiltonian $H$, can be developed in a power series as discussed by Magesan and Gambetta %(!!! more new notation !!!)
\cite{PhysRevA.101.052308} (cf. their Eq.~(A4)):

\begin{equation}
    X = e^{-i Z} = \mathbb{I} - i\lambda z_1 + \lambda^2 \left(-i z_2 - \frac{z_1^2 }{2}\right)+ \lambda^3\left(-i z_3 - \frac{1}{2}(z_1z_2+z_2z_1) +\frac{i}{6} z_1^3\right)+...
\end{equation}
where
\begin{equation}
    Z = \mathbb{I} + \lambda z_1 + \lambda^2 z_2 + ...
\end{equation}
introduces the generator operator $Z$. Previous literature does not have a consistent name for this operator; \cite{PhysRevA.101.052308} calls it $S$, only distinguishing it from the generator of $T$ by context.

The block diagonal part $\scaleobj{0.9}{\mathcal{B}}(X)$ is then given by 
\begin{multline}
   \scaleobj{0.9}{\mathcal{B}}(X) = \scaleobj{0.9}{\mathcal{B}}(e^{-i Z}) = \mathbb{I} - i\lambda \scaleobj{0.9}{\mathcal{B}}(z_1) + \lambda^2 \left(-i \scaleobj{0.9}{\mathcal{B}}(z_2) - \frac{\scaleobj{0.9}{\mathcal{B}}(z_1^2)}{2}\right)\\+ \lambda^3\left(-i \scaleobj{0.9}{\mathcal{B}}(z_3) - \frac{1}{2}(\scaleobj{0.9}{\mathcal{B}}(z_1z_2)+\scaleobj{0.9}{\mathcal{B}}(z_2z_1)) +\frac{i}{6}\scaleobj{0.9}{\mathcal{B}}(z_1^3)\right)+...
   \label{Baction}
\end{multline}

Putting equations (2) and (4) in (1) we get the power series of $T$ in terms of $z_1, z_2...$ etc. and using $T = e^{-i S}$ we can find a power series for generator $S$ in terms of $z_1,z_2,z_3...$. The details of the inistial steps of this derivation are given in Appendix \ref{appa}.

Let $T = \mathbb{I}+\lambda T_1 + \lambda^2 T_2...$.; on simplifying Eq.~(\ref{Tbig}) we get 
\begin{align}
\begin{aligned}
    T_1 = -i(z_1 - \scaleobj{0.9}{\mathcal{B}}(z_1));\\
    T_2 = -i(z_2 - \scaleobj{0.9}{\mathcal{B}}(z_2))-\frac{z_1^2}{2}-\frac{\scaleobj{0.9}{\mathcal{B}}(z_1)^2}{2}+z_1\scaleobj{0.9}{\mathcal{B}}(z_1);\\
    T_3 = -i(z_3 - \scaleobj{0.9}{\mathcal{B}}(z_3))+\frac{i}{6}(z_1^3 - \scaleobj{0.9}{\mathcal{B}}(z_1^3))-\frac{i}{2}\scaleobj{0.9}{\mathcal{B}}(z_1)^3-\frac{1}{2}(z_1z_2+z_2z_1) -\\
   \frac{1}{2}(\scaleobj{0.9}{\mathcal{B}}(z_1)\scaleobj{0.9}{\mathcal{B}}(z_2)+\scaleobj{0.9}{\mathcal{B}}(z_2)\scaleobj{0.9}{\mathcal{B}}(z_1)) +(z_1\scaleobj{0.9}{\mathcal{B}}(z_2)+z_2\scaleobj{0.9}{\mathcal{B}}(z_1))+\\
   \frac{i}{4}(\scaleobj{0.9}{\mathcal{B}}(z_1)\scaleobj{0.9}{\mathcal{B}}(z_1^2)+\scaleobj{0.9}{\mathcal{B}}(z_1^2)\scaleobj{0.9}{\mathcal{B}}(z_1)) -\frac{i}{2}(z_1^2 \scaleobj{0.9}{\mathcal{B}}(z_1) - z_1 \scaleobj{0.9}{\mathcal{B}}(z_1)^2).
\end{aligned}
\end{align}

Now $$T = e^{-i S} = \mathbb{I} - i\lambda s_1 + \lambda^2\left(-i s_2 -\frac{1}{2}s_1^2\right) +\lambda^3\left(-i s_3 -\frac{1}{2}(s_1s_2+s_2s_1)+\frac{i}{6}s_1^3\right)+...$$

Comparing coefficients we get 
\begin{align*}
    T_1 = -i s_1,\\
    T_2 = -i s_2 - \frac{s_1^2}{2},\\
    T_3 = -is_3 -\frac{1}{2}(s_1s_2+s_2s_1)+\frac{i}{6}s_1^3.
\end{align*}

On inverting the above relations we get

\begin{align}
    s_1 = i T_1,\\
    s_2 = i \left(T_2 + \frac{s_1^2}{2}\right),\\
    s_3 = i\left(T_3 +\frac{1}{2}(s_1s_2+s_2s_1)-\frac{i}{6}s_1^3\right).
\end{align}

Hence from equations (9), (10), (11) we can find the $s_1,s_2...$ in terms of $z_1,z_2,...$ recursively.

Up to $O(\lambda^3)$ we get (we begin here to use the notation [$a,b$] for the commutator of $a$ and $b$): 

%[\textbf{Note:} In the equations below [a,b] and \{a,b\} denote the commutator and anti-commutator between a and b respectively.]

\begin{align}
    s_1 = z_1 - \scaleobj{0.9}{\mathcal{B}}(z_1),\label{s1f}\\\nonumber\\
%    s_2 = (z_2 - \scaleobj{0.9}{\mathcal{B}}(z_2)) + \frac{i}{2}(z_1\scaleobj{0.9}{\mathcal{B}}(z_1)-\scaleobj{0.9}{\mathcal{B}}(z_1)z_1)\label{s1f}\nonumber\\
    s_2 = (z_2 - \scaleobj{0.9}{\mathcal{B}}(z_2)) + \frac{i}{2}[z_1,\scaleobj{0.9}{\mathcal{B}}(z_1)],\label{s2f}
\end{align}
\begin{multline}
    s_3 = (z_3 - \scaleobj{0.9}{\mathcal{B}}(z_3)) + \frac{\scaleobj{0.9}{\mathcal{B}}(z_1^3)}{6} +\frac{\scaleobj{0.9}{\mathcal{B}}(z_1)^3}{3} + \frac{i}{2}\left([z_1,\scaleobj{0.9}{\mathcal{B}}(z_2)] + [z_2,\scaleobj{0.9}{\mathcal{B}}(z_1)]\right)-\frac{1}{4}(
    \scaleobj{0.9}{\mathcal{B}}(z_1)\scaleobj{0.9}{\mathcal{B}}(z_1^2)+
     \scaleobj{0.9}{\mathcal{B}}(z_1^2)\scaleobj{0.9}{\mathcal{B}}(z_1))
    %\\ -\frac{1}{4}\{\scaleobj{0.9}{\mathcal{B}}(z_1),\scaleobj{0.9}{\mathcal{B}}(z_1^2)\} -\frac{1}{12}(\{\scaleobj{0.9}{\mathcal{B}}(z_1)^2,z_1\}-\{z_1^2,\scaleobj{0.9}{\mathcal{B}}(z_1)\}) - \frac{1}{6}(z_1\scaleobj{0.9}{\mathcal{B}}(z_1)z_1-\scaleobj{0.9}{\mathcal{B}}(z_1)z_1\scaleobj{0.9}{\mathcal{B}}(z_1))
    \\  -\frac{1}{12}(
    \scaleobj{0.9}{\mathcal{B}}(z_1)^2z_1+z_1\scaleobj{0.9}{\mathcal{B}}(z_1)^2
    -z_1^2\scaleobj{0.9}{\mathcal{B}}(z_1)-\scaleobj{0.9}{\mathcal{B}}(z_1)z_1^2) 
    - \frac{1}{6}(z_1\scaleobj{0.9}{\mathcal{B}}(z_1)z_1-\scaleobj{0.9}{\mathcal{B}}(z_1)z_1\scaleobj{0.9}{\mathcal{B}}(z_1)).\label{s3f}
\end{multline}

From this we finally get the series form of the block hamiltonian, the formula for which we simply transcribe for completeness from Eq.~(A5) of \cite{PhysRevA.101.052308}: 
\begin{multline}
    H_{block} = e^{iS}(H_0 + \lambda H_1)e^{-iS} \\= H_0 + \lambda(i[s_1,H_0]+H_1)+
    \lambda^2\left(i[s_2,H_0]-\frac{1}{2}[s_1,[s_1,H_0]] +i[s_1,H_1]\right) + \\
    \lambda^3\left(i[s_3,H_0] -\frac{i}{6}[s_1,[s_1,[s_1,H_0]]] -\frac{1}{2}([s_1,[s_2,H_0]] + [s_2,[s_1,H_0]])+i[s_2,H_1] -\frac{1}{2}[s_1,[s_1,H_1]]\right)\\+O(\lambda^4).
    \label{finaleq}
\end{multline}
% Couple of the subscripts of s in the above equation were incorrect/missing. I added it (straight from the Magesan paper).

%\fading{Also give for completeness the formula which goes from $s_i$ quantities to $H_{block}$ (basically, part of (A5) of Magesan) We should make a new notation for $(\cdot )_{BD}$. I recommend $\scaleobj{0.9}{\mathcal{B}}(\cdot )$:
%\begin{equation}
%  s_1 = z_1 - \scaleobj{0.9}{\mathcal{B}}(z_1)   
%\end{equation}
%\begin{equation}
%  s_1 = z_1 - (z_1)_{BD} 
%\end{equation}
% }

\section{Numerical Verification of Power series}
To verify the cumbersome algebra involved in obtaining the power series Eq.~(\ref{finaleq}) for effective block-diagonal Hamiltonian $H_{block}$, we use Mathematica and numerically compute the matrices and the series to various orders.\\
\href{https://drive.google.com/file/d/1lNNyNlPiuKJCSIl1sidpMupa0qJHbg1W/view?usp=sharing}{Click here for Mathematica notebook containing numerical verification}\\
We compare the obtained power series Eq.~(\ref{finaleq}) to the exact Cederbaum method Eq.~(\ref{ceder2}) and verify the scaling of errors (difference). The resulting block diagonal Hamiltonians $H_{block}$ as obtained from the power series and from the exact Cederbaum method \cite{L.S.Cederbaum_1989} differ only as $O(\lambda^4)$. Thus the obtained power series is indeed correct up to $O(\lambda^3)$.

\begin{figure}[H]
    \centering
    \includegraphics[width=1\linewidth]{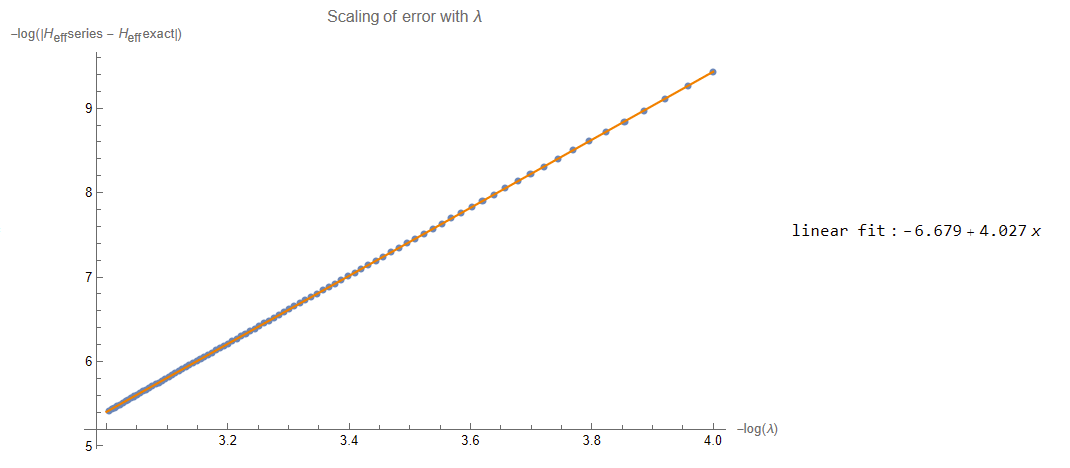}
    \caption{Numerical verification: Results match up to $O(\lambda^3)$}
    \label{fig:enter-label}
\end{figure}

\section{Conclusion and outlook}

In the above section we have found the perturbative series for the generator of the block diagonalising unitary $T$ up to $O(\lambda^3)$. We find that while the $s_1,s_2$ are completely block off-diagonal (i.e., $\scaleobj{0.9}{\mathcal{B}}(s_1)=\scaleobj{0.9}{\mathcal{B}}(s_2)=0$, see Eqs.~(\ref{s1f},\ref{s2f})), $s_3$ has non-zero block diagonal terms (i.e., $\scaleobj{0.9}{\mathcal{B}}(s_3)\neq 0$, see Eqs.~(\ref{s3f})). 

Our calculation has implications on application of the Cederbaum series to various scenarios. For example Magesan et al.\cite{PhysRevA.101.052308}, assumed the generator series $s_1,s_2...$ to be completely block-off diagonal to all orders (in their Appendix A.2.b). This assumption, which, as we have seen, is not compatible with the Cederbaum ``least action" principle, was then used in section IV A to find the effective Hamiltonian for superconducting qubits. We believe that the exact series should be used for accurate numerics in calculations. The assumption of block diagonal generator $S$ is only an approximation valid up to $O(\lambda^2)$ in the scale parameter $\lambda$. 

However, it seems that $\scaleobj{0.9}{\mathcal{B}}(S)=0$ is also a workable prescription for the block diagonalizing operator: and it has been successfully used in other works, and there is an efficient recursive procedure \cite{PhysRevA.101.052308} for generating terms in the series. We consider this superior (but see the alternative view of H{\"o}rmann and Schmidt \cite{10.21468/SciPostPhys.15.3.097} on this) to a series obtained by long-multiplication and inversion, as we have done for the ``least action" series. We would see as open questions two items: 1) Can the ``least action" series be generated in a recursive manner? 2) Can a resummation of the $\scaleobj{0.9}{\mathcal{B}}(S)=0$ formula give a manifestly analytic functional of $X$ as for the ``least action" case (Eq.~(\ref{ceder2}))?

%\section*{Acknowledgements}
%I (Ishan) would like to thank Prof. DiVincenzo for his valuable guidance and support throughout the course of this work.  
%\section*{Appendix}

\appendix

\setcounter{equation}{0}
\renewcommand{\theequation}{A.\arabic{equation}}

\section{Derivation of Power Series}
\label{appa}

Starting from Eqs.~(\ref{Hbasic}-\ref{Baction}) of the main text, we derive the power series of $T$ up to third order. All the equations written below are correct up to $O(\lambda^3)$.
 \begin{eqnarray}
   &\scaleobj{0.9}{\mathcal{B}}(X)\cdot&\!\!\!\!\!\scaleobj{0.9}{\mathcal{B}}(X^{\dagger})
   = \textcolor{red}{\scaleobj{0.9}{\mathcal{B}}(e^{-i Z})}\cdot\textcolor{blue}{\scaleobj{0.9}{\mathcal{B}}(e^{i Z})}  \\
  &= & \!\!\!\!\!\!\textcolor{red}{\left\{\mathbb{I} - i\lambda \scaleobj{0.9}{\mathcal{B}}(z_1) + \lambda^2 \left(-i \scaleobj{0.9}{\mathcal{B}}(z_2) - \frac{\scaleobj{0.9}{\mathcal{B}}(z_1^2 )}{2}\right)+ \lambda^3\left[-i \scaleobj{0.9}{\mathcal{B}}(z_3) - \frac{1}{2}(\scaleobj{0.9}{\mathcal{B}}(z_1z_2)+\scaleobj{0.9}{\mathcal{B}}(z_2z_1)) +\frac{i}{6} \scaleobj{0.9}{\mathcal{B}}(z_1^3)\right]\right\}}\cdot\nonumber\\ &&\!\!\!\!\!\!\textcolor{blue}{\left\{\mathbb{I} + i\lambda \scaleobj{0.9}{\mathcal{B}}(z_1) + \lambda^2 \left(i \scaleobj{0.9}{\mathcal{B}}(z_2) - \frac{\scaleobj{0.9}{\mathcal{B}}(z_1^2 )}{2}\right)+ \lambda^3\left[i \scaleobj{0.9}{\mathcal{B}}(z_3) - \frac{1}{2}(\scaleobj{0.9}{\mathcal{B}}(z_1z_2) +\scaleobj{0.9}{\mathcal{B}}(z_2z_1) ) -\frac{i}{6} \scaleobj{0.9}{\mathcal{B}}(z_1^3)\right]\right\}},\nonumber
\end{eqnarray}
which upon simplification gives,
\begin{multline}
      \scaleobj{0.9}{\mathcal{B}}(X)\cdot \scaleobj{0.9}{\mathcal{B}}(X^{\dagger}) = \mathbb{I} + \lambda^2 \left(-\scaleobj{0.9}{\mathcal{B}}(z_1^2) + \scaleobj{0.9}{\mathcal{B}}(z_1)^2\right) + \\ \lambda^3 \left\{ -(\scaleobj{0.9}{\mathcal{B}}(z_1z_2) + \scaleobj{0.9}{\mathcal{B}}(z_2z_1)) + (\scaleobj{0.9}{\mathcal{B}}(z_1)\scaleobj{0.9}{\mathcal{B}}(z_2) + \scaleobj{0.9}{\mathcal{B}}(z_2)\scaleobj{0.9}{\mathcal{B}}(z_1)) + \frac{i}{2}(\scaleobj{0.9}{\mathcal{B}}(z_1)\scaleobj{0.9}{\mathcal{B}}(z_1^2)-\scaleobj{0.9}{\mathcal{B}}(z_1^2)\scaleobj{0.9}{\mathcal{B}}(z_1))\right\}.
\end{multline}
Then up to $O(\lambda^3)$,
\begin{multline}
     \textcolor{brown}{\left\{\scaleobj{0.9}{\mathcal{B}}(X)\cdot\scaleobj{0.9}{\mathcal{B}}(X^{\dagger})\right\}^{-1/2} = \mathbb{I} - \frac{\lambda^2}{2} (-\scaleobj{0.9}{\mathcal{B}}(z_1^2) + \scaleobj{0.9}{\mathcal{B}}(z_1)^2) -}\\ \textcolor{brown}{\frac{\lambda^3}{2} \left\{ -(\scaleobj{0.9}{\mathcal{B}}(z_1z_2) + \scaleobj{0.9}{\mathcal{B}}(z_2z_1)) + (\scaleobj{0.9}{\mathcal{B}}(z_1)\scaleobj{0.9}{\mathcal{B}}(z_2) + \scaleobj{0.9}{\mathcal{B}}(z_2)\scaleobj{0.9}{\mathcal{B}}(z_1)) + \frac{i}{2}(\scaleobj{0.9}{\mathcal{B}}(z_1)\scaleobj{0.9}{\mathcal{B}}(z_1^2)-\scaleobj{0.9}{\mathcal{B}}(z_1^2)\scaleobj{0.9}{\mathcal{B}}(z_1))\right\}}
\end{multline}
Similarly, 
\begin{multline}
   X\cdot\scaleobj{0.9}{\mathcal{B}}(X^{\dagger})
   = \textcolor{red}{(e^{-i Z})}\cdot\textcolor{blue}{\scaleobj{0.9}{\mathcal{B}}(e^{i Z})}
  = \textcolor{red}{\left\{\mathbb{I} - i\lambda z_1 + \lambda^2 \left(-i z_2 - \frac{z_1^2 }{2}\right)+ \lambda^3\left[-i z_3 - \frac{1}{2}(z_1z_2+z_2z_1) +\frac{i}{6} z_1^3\right]\right\}}\cdot\\ \textcolor{blue}{ \left\{\mathbb{I} + i\lambda \scaleobj{0.9}{\mathcal{B}}(z_1) + \lambda^2 \left(i \scaleobj{0.9}{\mathcal{B}}(z_2) - \frac{\scaleobj{0.9}{\mathcal{B}}(z_1^2 ) }{2}\right)+ \lambda^3\left[i \scaleobj{0.9}{\mathcal{B}}(z_3) - \frac{1}{2}(\scaleobj{0.9}{\mathcal{B}}(z_1z_2) +\scaleobj{0.9}{\mathcal{B}}(z_2z_1) ) -\frac{i}{6} \scaleobj{0.9}{\mathcal{B}}(z_1^3)\right]\right\}}
\end{multline}
Which on simplification gives
\begin{multline}
   \textcolor{orange}{X\cdot\scaleobj{0.9}{\mathcal{B}}(X^{\dagger})
  = \mathbb{I} -i\lambda(z_1 - \scaleobj{0.9}{\mathcal{B}}(z_1)) + \lambda^2 \left\{-i(z_2 - \scaleobj{0.9}{\mathcal{B}}(z_2)) - \frac{z_1^2}{2} - \frac{\scaleobj{0.9}{\mathcal{B}}(z_1^2)}{2} + z_1 \scaleobj{0.9}{\mathcal{B}}(z_1)\right\} + }\\ \textcolor{orange}{\lambda^3\Bigg\{ \Bigg .-i(z_3-\scaleobj{0.9}{\mathcal{B}}(z_3))-\frac{1}{2}(\scaleobj{0.9}{\mathcal{B}}(z_1z_2)+\scaleobj{0.9}{\mathcal{B}}(z_2z_1))-\frac{1}{2}(z_1z_2+z_2z_1)+}\\ \textcolor{orange}{\frac{i}{6}(z_1^3 - \scaleobj{0.9}{\mathcal{B}}(z_1^3))+(z_1\scaleobj{0.9}{\mathcal{B}}(z_2)+z_2\scaleobj{0.9}{\mathcal{B}}(z_1))
  +\frac{i}{2}(z_1\scaleobj{0.9}{\mathcal{B}}(z_1^2) - z_1^2\scaleobj{0.9}{\mathcal{B}}(z_1))\Bigg.\Bigg\}}.
\end{multline}
From equation (1) $T = \textcolor{orange}{X\cdot\scaleobj{0.9}{\mathcal{B}}(X^{\dagger})}\textcolor{brown}{\cdot(\scaleobj{0.9}{\mathcal{B}}(X) \scaleobj{0.9}{\mathcal{B}}(X^{\dagger}))^{-1/2}}$; using equations (5) and (6) we get 
\begin{multline}
    T = \textcolor{orange}{\Bigg\{ \Bigg.\mathbb{I} -i\lambda(z_1 - \scaleobj{0.9}{\mathcal{B}}(z_1)) + \lambda^2\left(-i(z_2 - \scaleobj{0.9}{\mathcal{B}}(z_2)) - \frac{z_1^2}{2} - \frac{\scaleobj{0.9}{\mathcal{B}}(z_1^2)}{2} + z_1 \scaleobj{0.9}{\mathcal{B}}(z_1)\right) + }\\ \textcolor{orange}{\lambda^3\Bigg( \Bigg.-i(z_3-\scaleobj{0.9}{\mathcal{B}}(z_3))-\frac{1}{2}(\scaleobj{0.9}{\mathcal{B}}(z_1z_2)+\scaleobj{0.9}{\mathcal{B}}(z_2z_1))-\frac{1}{2}(z_1z_2+z_2z_1)+}\\ \textcolor{orange}{\frac{i}{6}(z_1^3 - \scaleobj{0.9}{\mathcal{B}}(z_1^3))+(z_1\scaleobj{0.9}{\mathcal{B}}(z_2)+z_2\scaleobj{0.9}{\mathcal{B}}(z_1))
  +\frac{i}{2}(z_1\scaleobj{0.9}{\mathcal{B}}(z_1^2) - z_1^2\scaleobj{0.9}{\mathcal{B}}(z_1))\Bigg.\Bigg)\Bigg.\Bigg\}}.
 \textcolor{brown}{\Bigg\{\Bigg.\mathbb{I} - \frac{\lambda^2}{2} (-\scaleobj{0.9}{\mathcal{B}}(z_1^2) + \scaleobj{0.9}{\mathcal{B}}(z_1)^2) - }\\ \textcolor{brown}{\frac{\lambda^3}{2} \Bigg( -(\scaleobj{0.9}{\mathcal{B}}(z_1z_2) + \scaleobj{0.9}{\mathcal{B}}(z_2z_1)) + (\scaleobj{0.9}{\mathcal{B}}(z_1)\scaleobj{0.9}{\mathcal{B}}(z_2) + \scaleobj{0.9}{\mathcal{B}}(z_2)\scaleobj{0.9}{\mathcal{B}}(z_1)) + \frac{i}{2}(\scaleobj{0.9}{\mathcal{B}}(z_1)\scaleobj{0.9}{\mathcal{B}}(z_1^2)-\scaleobj{0.9}{\mathcal{B}}(z_1^2)\scaleobj{0.9}{\mathcal{B}}(z_1))\Bigg)\Bigg.\Bigg\}}
 \label{Tbig}
\end{multline}

\bibliography{ref}

\end{document}